# VO: A Strongly Correlated Metal Close to a Mott-Hubbard Transition.

F. Rivadulla[1*], J. Fernández-Rossier[2], M. García-Hernández[3], M. A. López-Quintela[1], J. Rivas[4], J. B. Goodenough[5].

[1]*Physical-Chemistry Department, University of Santiago de Compostela, 15782-Santiago de Compostela, Spain.*

[2] *Applied Physics Department, University of Alicante, San Vicente del Raspeig 03690, Alicante, Spain.*

[3]*Instituto de Ciencia de Materiales de Madrid (CSIC), Cantoblanco, E-28049 Madrid, Spain*

[4]*Applied Physics Department, University of Santiago de Compostela, 15782-Santiago de Compostela, Spain.*

[5]*Texas Materials Institute, ETC 9.102, The University of Texas at Austin, 1 University Station, C2201, Austin, Texas 78712, USA*

*e-mail: qffran@usc.es

**Here we present experimental and computational evidences to support that rock-salt cubic VO is a strongly correlated metal with Non-Fermi-Liquid thermodynamics and an unusually strong spin-lattice coupling. An unexpected change of sign of metallic thermopower with composition is tentatively ascribed to the presence of a pseudogap in the density of states. These properties are discussed as signatures of the proximity to a magnetic quantum phase transition. The results are summarized in a new electronic phase diagram for the 3d monoxides, which resembles that of other strongly correlated systems. The structural and electronic simplicity of 3d monoxides make them ideal candidates to progress in the understanding of highly correlated electron systems.**

## INTRODUCTION

The outer d electrons of most transition metals occupy narrow bands that, as long as the bandwidth (W) remains larger than the intraatomic coulombic repulsion (U), retain an itinerant-electron character described by the Landau Fermi-Liquid (FL) picture. The FL theory correctly predicts that low temperature magnetic susceptibility, $\chi$, specific heat *C(T)*, and resistivity $\rho(T)$ scale as $T^0$, $T^1$ and $T^2$ respectively.[1]

However, for materials with U>W the interatomic interactions are not strong enough to screen effectively the intraatomic interactions so that electrons become localized and the spectrum of charged excitations acquires a gap. These are the so called Mott-Hubbard insulators[2]. It has long been recognized that the collective quantum states of conductors and insulators are fundamentally different phases separated by some kind of quantum phase transition (QPT) boundary. In the neighborhood of the QPT, interactions are non-negligible and $\chi$, *C/T* and $\rho(T)$ deviate from the FL scenario. This behavior has been observed in heavy fermions[3,4], cuprates[5], manganites[6] and even simple metallic alloys[7] and other materials[8] close to a magnetic QPT. Moreover, a depression of the density of states around the Fermi Energy, the so called pseudo-gap, has been shown to occurr in many of these systems.[9,10,11]

In rock-salt transition metal (TM) monoxides (TiO, VO, MnO, FeO, CoO and NiO), octahedral-site M-O interactions split the 3d orbitals into a more stable, threefold-degenerate manifold of $\pi$-bonding $t_{2g}$ orbitals (xy, yz $\pm$ izx) and a twofold-degenerate manifold of $\sigma$-bonding $e_g$ orbitals by an energy $\Delta_c$. Occupation of these bands determine the electronic properties across the series: correlation driven insulators with antiferromagnetic order in the case of MnO, FeO, CoO and NiO, and a Fermi-liquid metal, with a superconducting phase below 1K, in the case of TiO.[12] Naively, a quantum phase transition that separates metallic TiO from the insulating antiferromagnet MnO can be accomplished as additional electrons are added into the *d* levels of isoestructural TM monoxides. In the crossover region between these two antagonistic phases are located CrO, whose bulk synthesis remains a challenge, and VO whose electronic properties are discussed here.

In this paper we report experimental measurements of specific heat, electronic transport and magnetic susceptibility of a series of samples of $TiO_x$ and $VO_x$, with 0.9<x<1.1. In spite of their similar electronic structure, we will present solid evidence that TiO is well described by conventional FL band theory, while VO departs from FL picture. The results are discussed in terms of the proximity to a magnetic/electronic quantum phase transition.

## EXPERIMENTAL

Polycrystalline $VO_x$ and $TiO_x$ have been generally synthesized by arc melting and casting. This method presents important experimental difficulties that have resulted in disagreement between data published before the work of Banus, Reed and Strauss.[13] To avoid these problems we propose an alternative synthetic route that yields $VO_X$ with controllable stoichiometry and of quality comparable to traditional methods. $TiO_x$ and $VO_x$ are perfectly stable in a wide compositional range, approximately between $0.8<x<1.2$,[13] and even the stoichiometric compounds (x=1) show 16% of vacancies at both the metal and oxygen sites to shorten the lattice parameter so as to increase W.[14] For the synthesis of $VO_x$, high purity vanadium metal and $V_2O_3$ were mixed in stoichiometric proportions, according to the desired value of x. The powders were ground, mixed, and pressed into pellets in an Ar atmosphere before being transferred and sealed into a silica tube that had been evacuated down to $P \approx 10^{-5}$-$10^{-4}$ Torr. The pellet was placed in a small alumina crucible to avoid reaction of V with the tube, which produces traces of $V_3Si$, difficult to detect by x-ray. The ampoules were annealed at 1100ºC for 24 h and quenched into an ice-water bath. Quenching from high temperature avoids problems of disproportionation, which was probably the origin of the metal-insulator transition attributed to VO in the oldest literature.[15] Quenched pellets were ground and cold-pressed at $16\times10^3$ kg/cm$^2$ before being again sealed in evacuated silica tubes and refired at 1100ºC for 24 h. After this treatment the pellets (shining-grey) were polished and the x-ray patterns showed only very narrow peaks of the single-phase cubic (Fm-3m) material (Fig.1). The oxygen/metal ratio x in $VO_x$ was determined by thermogravimetric analysis, TGA (Figure 1). Powdered samples were calcined in oxygen at 1ºC/min up to 650ºC and held for 16 h. After this time, oxidation to $V_2O_5$ was complete; no further weight gain was detected, indicating complete combustion of the monoxide. The lattice parameters for different x are in perfect agreement with the literature values[13] (Figure 1). Moreover, due to the correlation between lattice parameter and vacancy concentration, we can ensure a similar amount of vacancies as that reported in reference [13].The grain sizes in the sintered pellets determined by optical microscopy typically ranged between 5 and 20 microns.

Cubic $TiO_x$ (shining-gold) can be synthesized in a similar way from Ti and $TiO_2$ at higher temperature. TiO is commercially available and no difference has been observed between our samples and those purchased from Alfa. Annealing of TiO pellets at 1100ºC for 24 h under vacuum and slow cooling (holding the sample at 900ºC for at least 48 h.) results in a monoclinic phase (space group *A2/m*) with an ordered array of vacant lattice sites: half of the Ti and half of the O atoms are missing alternately in every third (110) plane. This process is completely reversible and both ordered/dissordered samples are stoichiometrically identical, within the error. All attempts to order the vacancies in the case of $VO_x$ were unsuccessful.

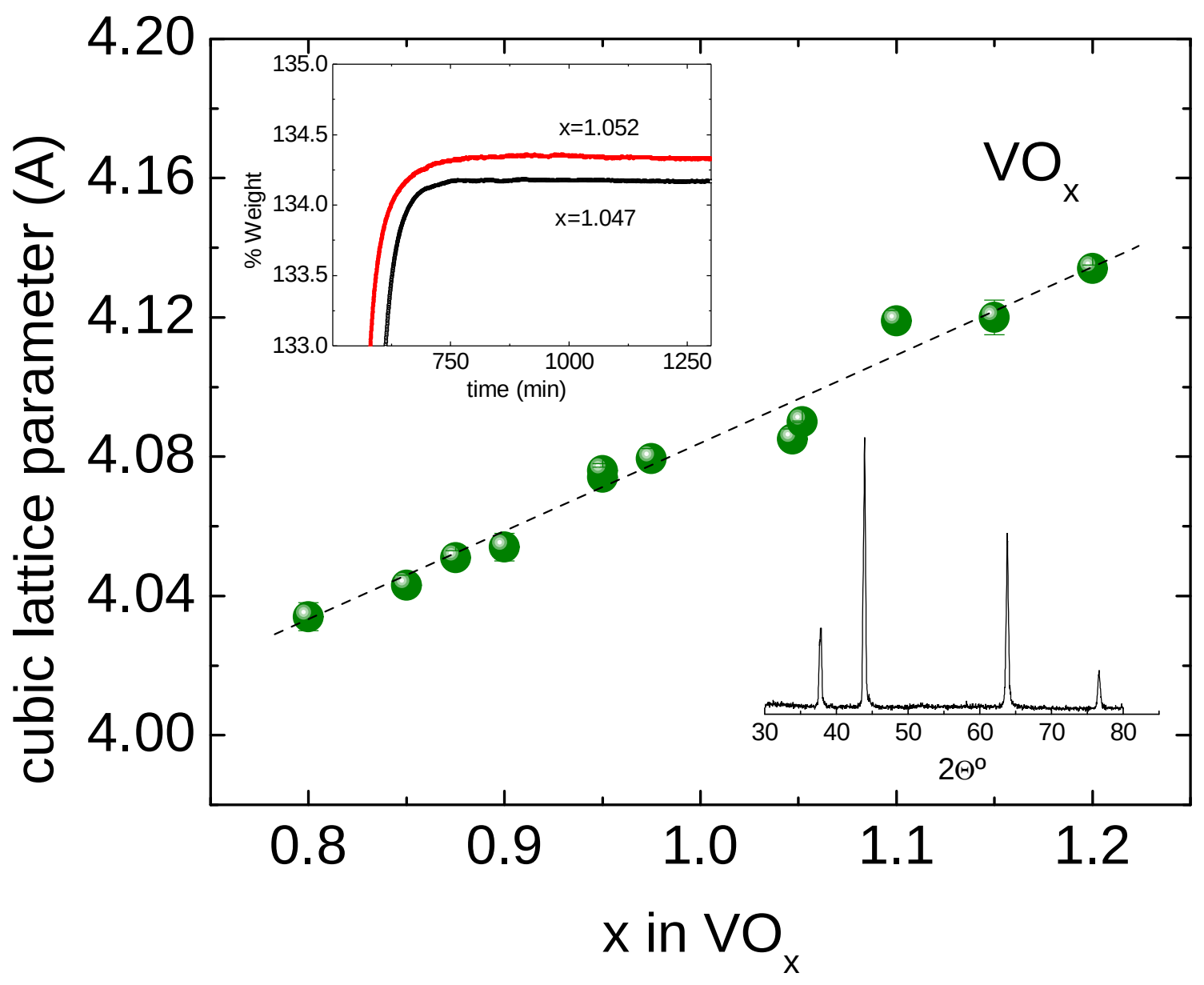

**Figure 1.** Evolution of lattice parameters in $VO_x$ with x. The results are practically identical to those obtained by Banus, Reed and Strauss (ref. 13), whose samples where synthesized by arc-melting and casting. Lower inset: example of x-ray pattern, in this case for $VO_{1.05}$. Upper inset: Detail of the final part of a TGA experiment to determine the oxygen content. Accuracy of this method is better than 0.01, as it can be observed.

## RESULTS AND DISCUSSION

Figures 2a to 2f show the energy bands and density of states of TiO, VO and hypothetical CrO, calculated with a gradient corrected local density approximation and localized atomic orbitals, using the CRYSTAL03[16]. The calculations shown in the figure assume spin unpolarized solutions and the lattice parameter that minimizes the ground state energy: 4.21Å, 4.16 Å and 4.11 Å for TiO, VO and CrO respectively. Under these approximations, the electronic structure of the three compounds is similar. They are metallic, with the *d* bands well separated from the s bands. In the case of TiO the Fermi energy lies well in the $t_{2g}$ bands, which in all the three compounds overlap with the $e_g$ bands. Moving from TiO to CrO the bandwidth decreases, and the Fermi energy moves upwards, to accommodate new electrons in states with some weight in the $e_g$ bands. Similar results are obtained using either B3LYP hybrid density functional[17] or the GW approach[18].

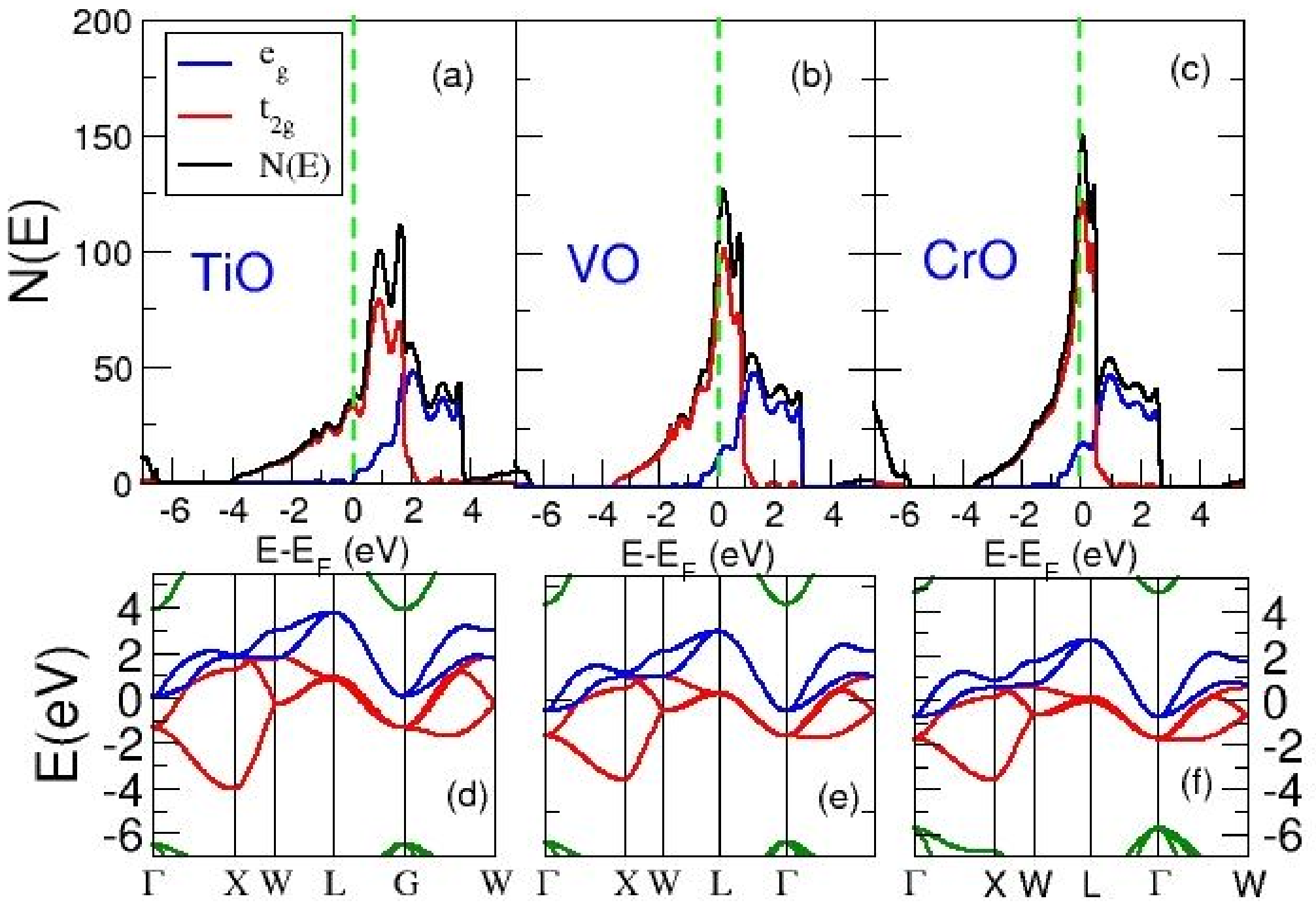

**Figure 2.** Electronic structure of TiO (left), VO (middle) and CrO (right) using density functional calculations in the gradient corrected approximation. In panels (a), (b) and (c) we show the total density of states as well as the projections to the $t_{2g}$ and $e_g$ levels (red and blue lines respectively). The vertical line shows the Fermi energy. In panels (d), (e) and (f) we show the bands closer to the Fermi Energy. The *d* $t_{2g}$ and $e_g$ bands are well separated from the *sp* bands.

The first experimental evidence of unconventional metallic behaviour in VO comes from magnetic susceptibility and specific heat measurements of both $VO_{0.95}$ and $TiO_{1.0}$, with and without ordered vacancies. These compositions are representative of the behaviour found for other values of x. For $TiO_{1.0}$, a temperature independent $\chi$(T) (Fig. 3) and the asymptotic evolution of C(T)/T towards a constant low temperature value (inset of Fig. 3) are nicely consistent with the expectations of a standard Fermi Liquid, independent of disorder. In contrast the susceptibility of $VO_X$ is strongly enhanced upon cooling below ≈20K, and follows a power law $(\chi-\chi_0) \approx T^{-\eta}$, with $\chi_0 \approx 2.7\times10^{-4}$ emu $mol^{-1}Oe^{-1}$, and $\eta \approx 0.78$ below ≈20 K (Fig. 3); a system of independent spins would obey a Curie law, with $\chi_0 \approx 0$ and $\eta \approx 1$. Application of a magnetic field flattens the $\chi$(T) curve, consistent with the opening of a Zeeman spin gap and the subsequent depletion of $\chi$(T).

Specific heat at low temperature in $VO_{0.95}$ also deviates from the linear temperature dependence and shows a strong magnetic field dependence (inset of Fig. 3), in both cases in conflict with Fermi-liquid theory. C(T)/T in $VO_X$ follows a power law $\propto T^{-\eta}$, $\eta \approx 0.65$, bellow ≈ 10

K. Moreover, it is evident a departure from this behaviour below ≈2K, due to the existence of a broad maximum, which becomes evident after application of a magnetic field.

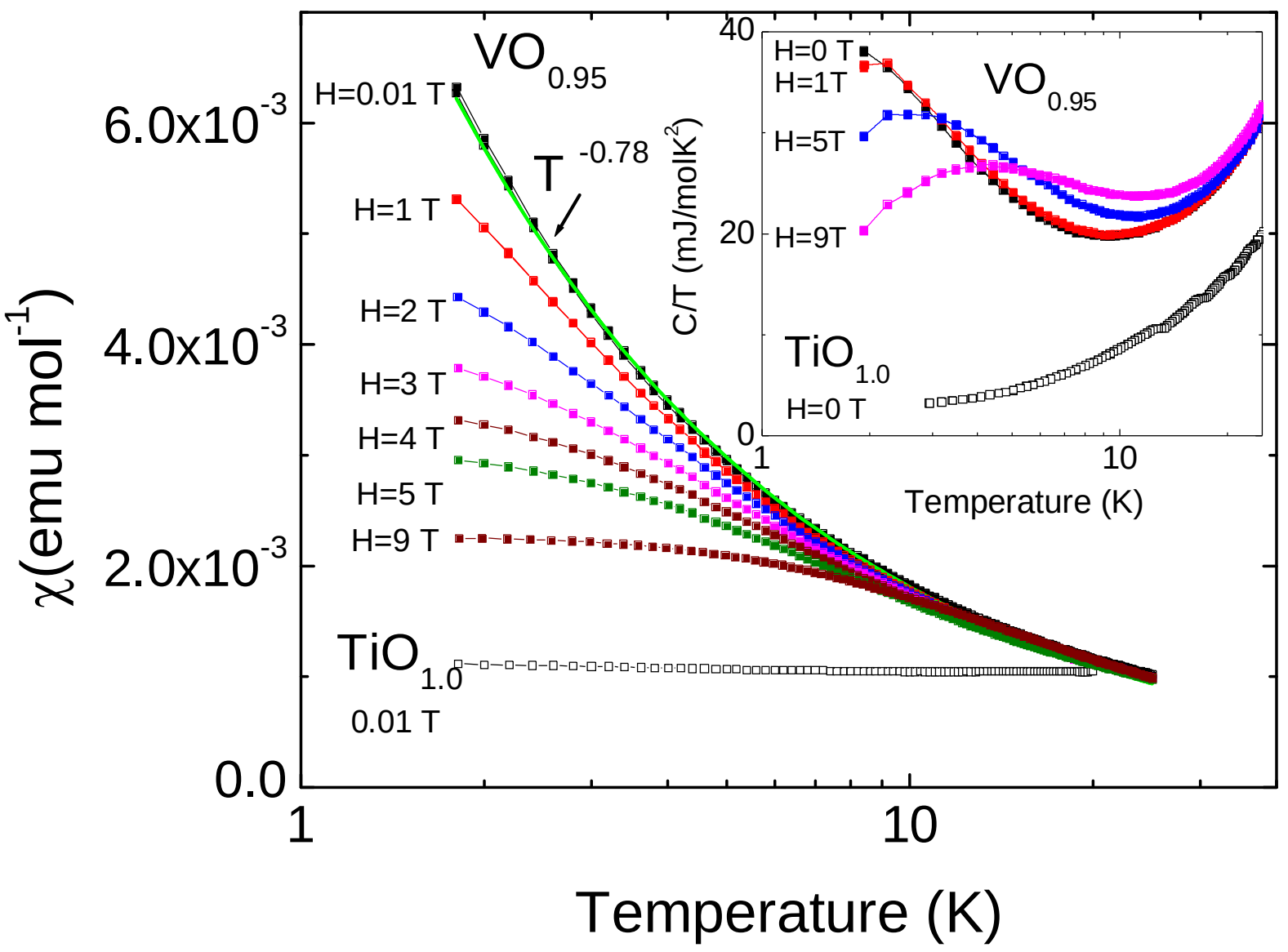

**Figure 3.** Low temperature susceptibility of $VO_{0.95}$ and $TiO_{1.0}$ at different fields. The behaviour is representative of that observed for other values of x. The green line is a fit to $(\chi - \chi_0) \propto T^{-0.78}$ Inset: C/T versus T for the same samples, at different magnetic fields.

The behaviour of the magnetic susceptibility and the specific heat in VO is absolutely unexpected and at first sight could remind that of a spin-glass above the freezing temperature. This would imply the presence of (interacting) localized spins in the FL, which is, a priori, unexpected for a conventional metal. In fact, a magnetic phase transition at T=0 plus disorder could lead to the observed spin-glass like features in the susceptibility and specific heat.[19] So, departure of the susceptibility and specific heat from the standard behaviour could be due to the tendency of d electrons to form local spin moments in VO, resulting in some kind of collective relaxation state.[20] The possibility of field-tuning the properties of this system makes it very interesting from an experimental point of view. Following this argument, the collective state reminiscent of a spin-glass would be an effect of the underlying electronic mechanism that produces the observed non-FL.

A priori, the origin of local magnetic moments and hence the failure of standard band theory to describe VO could be due either to disorder or to strong electronic correlations due to $W \approx U$. Disorder is certainly present in these compounds, which show a large number of vacancies at both the metal and oxygen sites. However, TiO has the same amount of vacancies but do not show any signature compatible with the presence of local spin moments.

On the other hand, we have not observed any correlation between the number of vacancies present at different compositions and the apparent divergence of the $\chi(T)$ and C(T)/T.

Charged elementary excitations are probed in transport experiments. In the inset of figure 4 we show resistivity versus temperature ρ*(T)* for the same sample of TiO before and after ordering the vacancies. It is apparent that the disordered cubic crystal behaves like a semiconductor whereas the ordered monoclinic crystal has a metallic like conductivity, with a large temperature independent part due to vacancy scattering. Therefore, ρ*(T)* curves are dominated by disorder and do not provide straightforward information about the effect of interactions on the quasiparticle dynamics.

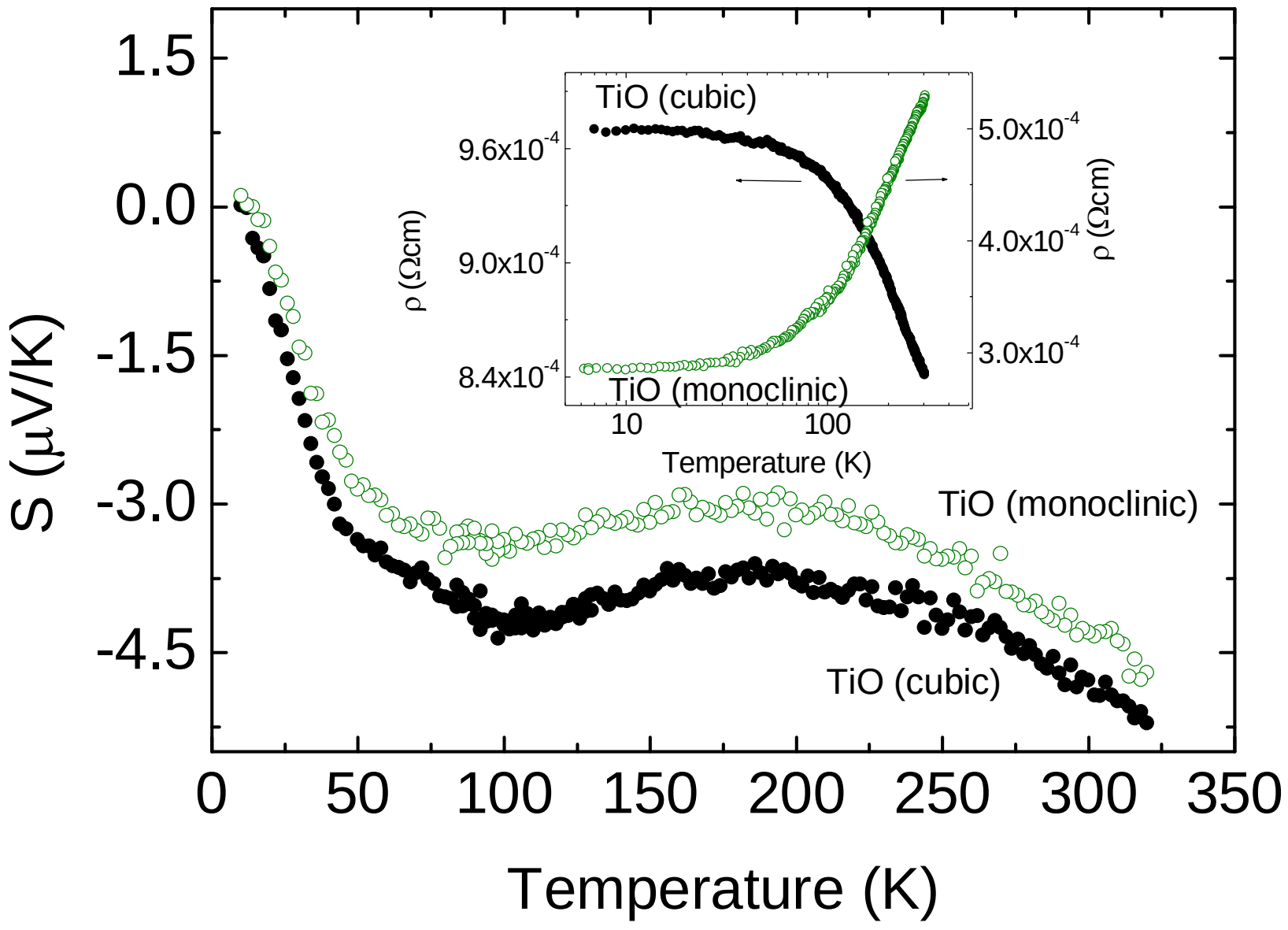

**Figure 4.** Thermoelectric power of $TiO_{1.0}$ with the vacancies ordered (monoclinic) and disordered (cubic). The small difference in the absolute value (≤0.5μV/K) must be due to a small variation in the stoichiometry produced by the annealing. In any case the difference is irrelevant and the temperature dependence is clearly not affected by the ordering of the vacancies. Inset: effect of vacancy ordering on the resistivity of $TiO_{1.0}$.

Then, the observed dρ/dT<0 in $VO_x$, is not intrinsic, but dominated by vacancy scattering.

In contrast, we observe that thermoelectric power in TiO is not sensitive to disorder (Fig. 4). At low temperature, the phonon-drag enhancement dominates over the contribution from conventional electronic diffusion. Close to room temperature the electronic contribution of itinerant charge carriers to the thermoelectric power, both in the case of conventional[21] and strongly correlated metals[22] is given by:

$$S = -C\frac{k_B}{e}\left(\frac{k_B T}{Z}\left.\frac{d\ln\Phi(E)}{dE}\right|_{E=E_F}\right) \quad (1)$$

where C is a dimensionless constant, e is the charge of the electron, $k_B$ is the Boltzmann constant, T is the temperature and $\Phi(E) = \frac{1}{V}\sum_k \left(\frac{\partial \varepsilon_k}{\partial k_x}\right)^2 \delta(E-\varepsilon_k)$ is a transport function with a energy dependence similar to that of the density of states *N(E)*. The thermoelectric power of a metal expressed by eq.(1) has an intrinsic sign, which reflects the curvature of *N(E)* around $E_F$. The negative S(T) in $TiO_X$ irrespective of x,[13] is consistent with a one-third filled $\pi$*-band, as obtained in the calculations. In contrast, the gradual change in the sign of S(T) with x in $VO_X$ (see Fig. 5) signals a change in the curvature of the density of states around $E_F$, as it crosses the mid-band energy, which is at odds with equation (1). In fact, the change in the sign of S(T) reported in Fig. 5 is compatible with a depression in the density of states (a pseudogap) around the Fermi energy in $VO_x$.

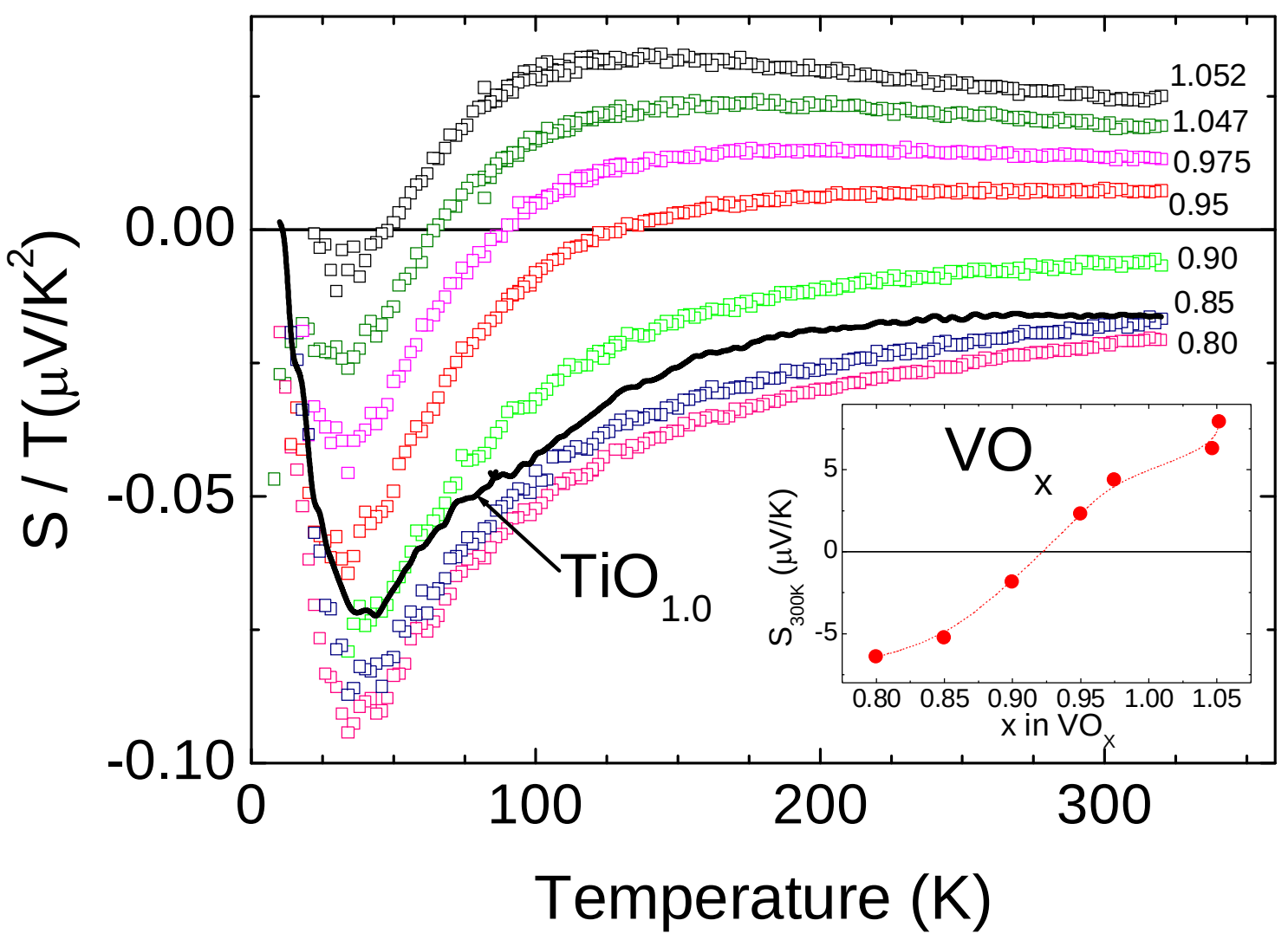

**Figure 5.** Thermoelectric power divided by temperature as a function of temperature. According to expression (1), the plot should be a constant value for a metal, at high enough temperature. At low temperatures, the phonon-drag enhancement deviates the experimental values from the diffusion formula. Inset: Variation of the thermopower at 300 K with x in $VO_X$. The change of sign occurs at x~0.95, consistent with X-ray absorption spectroscopy results [ref. 23], that show a valence change $(2-\delta)^+ \rightarrow (2+\delta)^+$ at x~0.94-0.97.

Having more than one band that crosses the Fermi Energy should not change this interpretation of the thermopower, as the dominant mechanism will be the scattering of charge carriers from the wider to the narrower bands (due to the higher density of states), introducing an scattering rate proportional to N(E). As a result, the thermopower will present a term which is proportional to the derivative of the density of states of the narrower band with respect to the energy, at $E_F$.

It is important to note that the change in the chemical potential necessary to account for the electron density difference between $VO_{0.9}$ and $VO_{1.1}$ is much smaller than the typical energy scales in which the *ab-initio* density of states varies. Therefore, the change of sign of the thermopower is due to dynamical electronic correlations absent in DFT calculations. In particular, short-range spin correlations in a doped Mott insulator can give rise to a pseudogap in the density of states[24]. These short-range spin correlations would be responsible also for the anomalous temperature dependence in $\chi$(T), and the magnetic field dependent of C(T)/T. Another indication of the connection between the anomalous magnetic and transport properties of $VO_x$ comes from the high *positive* magnetoresistance observed by Rata *et al.*[25]

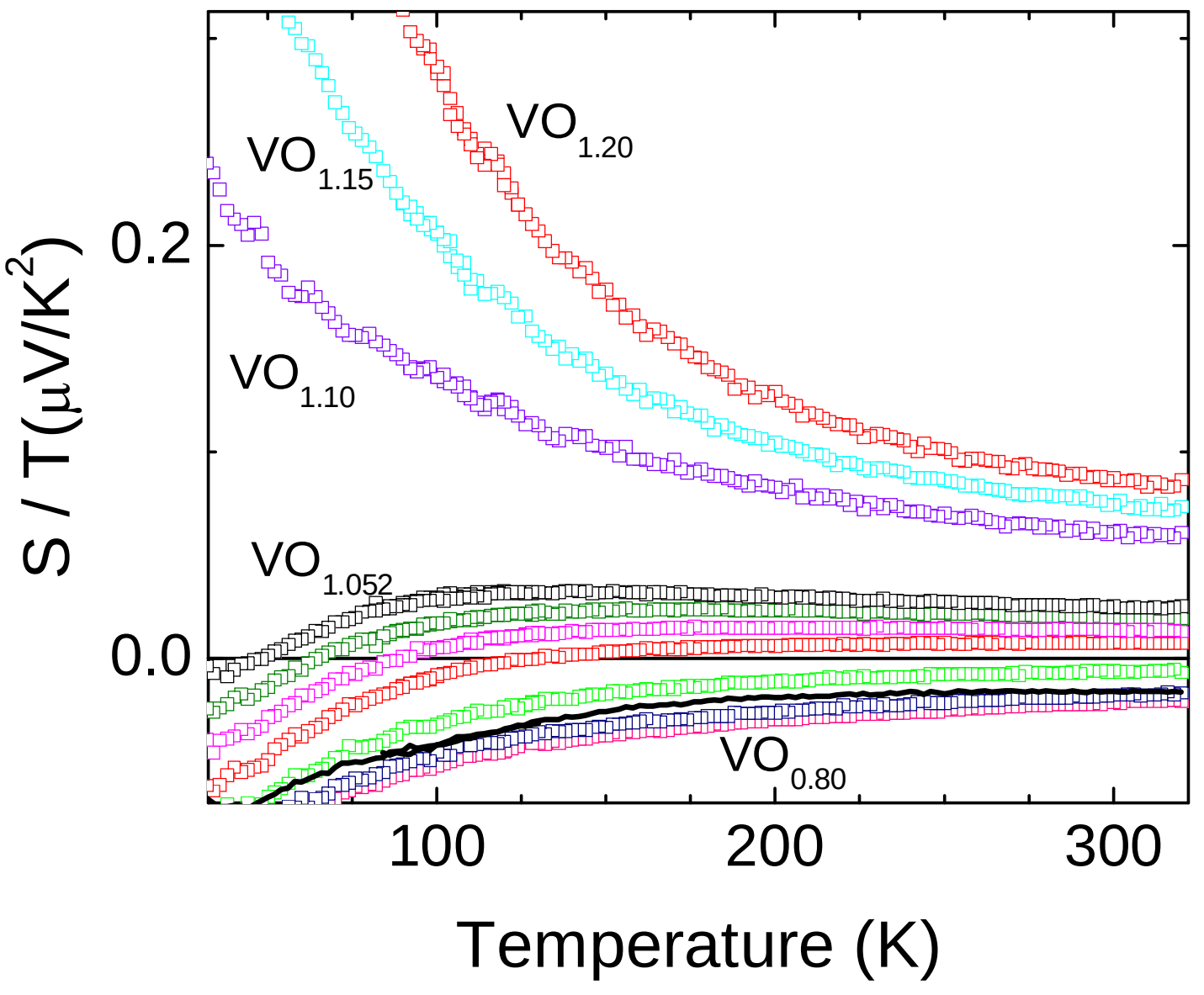

**Figure 6.** Themoelectric power vs. temperature for samples with a large amount of vacancies. Random quenched disorder introduces localized states and departure of intrinsic metallic behaviour.

One possibility against the correlation-driven pseudogap is the appearance of a mobility gap due to Anderson localization, which could give rise to local moments due to disorder. Random distribution of both $V^{2+}$ and $O^{2-}$ vacancies in $VO_x$ introduces a variation in the periodic potential from site to site, that could localize the electronic wave-functions if strong enough. This is expected to occur above a critical value of the ratio between the random potential and the bandwidth. From the band structure calculations performed in Fig.2, the difference in the bandwidth between TiO and VO is very small so that the effect of disorder in both materials must be very similar. The negligible effect in TiO almost completely rules out this explanation.
However, to fully discard Anderson localization, we have synthesized samples of $VO_x$ approaching the limit of solubility of the system ($x \approx 0.2$) and measured their thermopower. As x increases in $VO_x$, the number of V vacancies increases and hence, the perturbation of the periodic potential experienced by the conduction electrons. For samples with a large amount of V vacancies, the thermopower deviates from the Mott formula and shows clear signs of activated behaviour (Fig. 6). However, disordered vacancies in VO do not put $E_F$ bellow a mobility edge for the range $0.8<x<1.1$, studied in this work.

Our experimental results demonstrate that the low energy elementary excitation spectrum of VO is dominated by some kind of spin fluctuation without long range order. The tendency of VO to develop local magnetic moments is supported by our *spin polarized* density functional calculations. For VO and CrO, either ferromagnetic or antiferromagnetic solutions are much lower in energy than the paramagnetic one, both within the GGA and B3LYP functionals. In contrast, the paramagnetic electronic structure of TiO calculated with the GGA functional has a smaller energy than the spin polarized solutions, in agreement with the experiment. This is another indication that whereas TiO is a band conductor, VO and CrO have a tendency to develop local moments with spin 3/2 and 4/2 respectively.

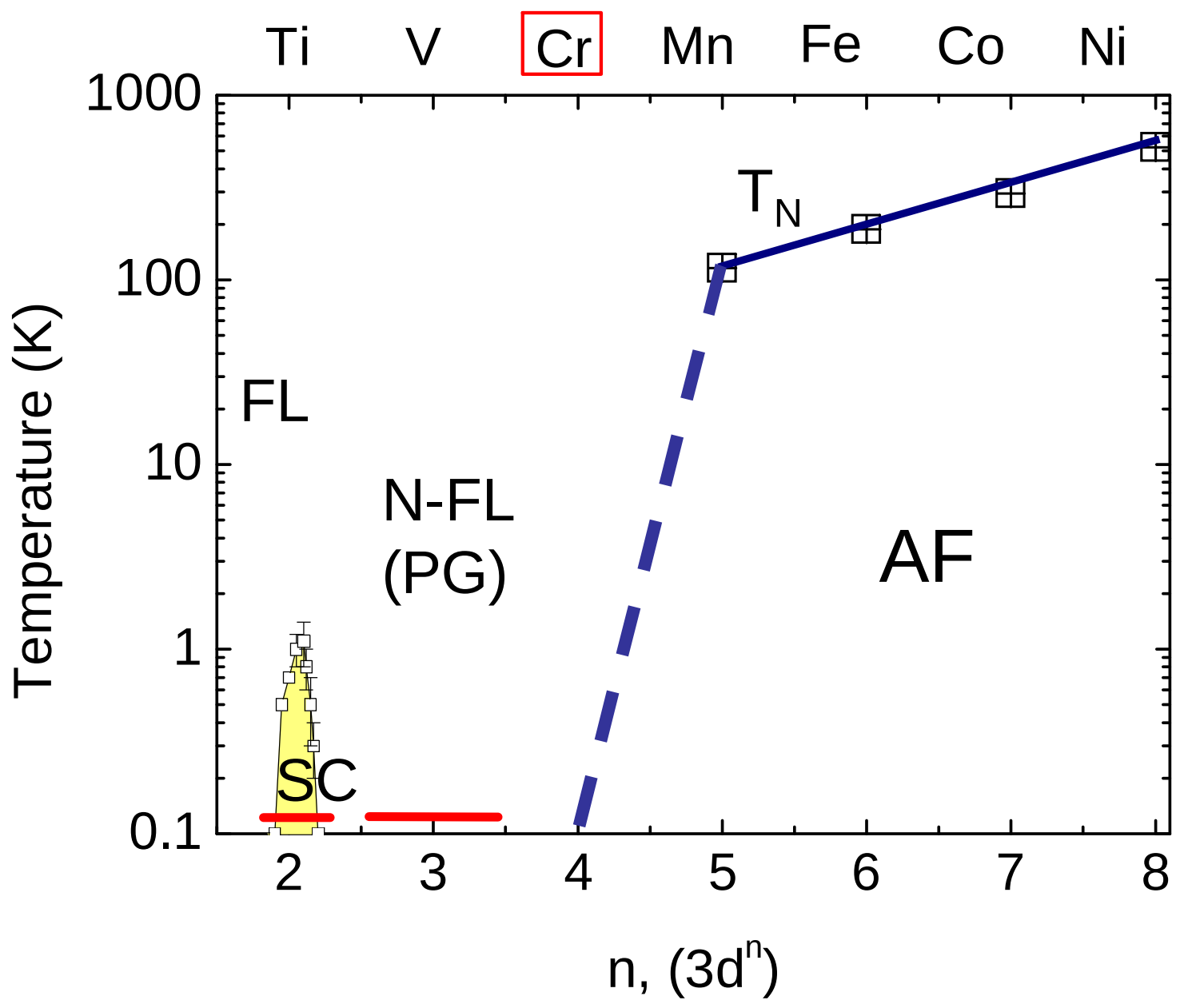

**Figure 7.** Unified electronic phase diagram of the monoxides of the first transition series. It does represent a generic electronic/magnetic phase diagram of a cubic (rock-salt) structure, in which 3d electrons are successively added to it. In spite of the simplicity of the structure and its three dimensional character, the diagram shows many of the peculiarities of the diagram of more complex systems. The red horizontal bars indicate the range of doping explored in $TiO_X$ and $VO_X$. Cr is squared to signal that bulk CrO does not exist. The the dotted line represents the uncertainty to locate precisely the quantum phase transition. FL stands for Fermi-Liquid; N-FL, non-Fermi Liquid; PG, pseudogap; SC, superconductor; AF, antiferromagnet; and $T_N$ is the Neel Temperature.

The development of local moments would have consequences on the stability of the lattice. According to the electronic structure calculations, the equilibrium lattice constant of VO is 4.3 for spin polarized solutions and 4.16 for spin unpolarized ones. Thereby, the spin fluctuations revealed by our experiments would be accompanied by strong lattice fluctuations. This scenario of bond-length fluctuations corroborates that proposed earlier[26] to account for the suppression of the phonon contribution to the thermal conductivity and of vacancy ordering. This spin-lattice coupling is also obtained in our calculations for CrO. In this case the occupation of the orbitals depends on the spin: whereas paramagnetic CrO accommodates 4 electrons mostly in the $t_{2g}$ bands, our density functional calculations, both in with GGA and B3LYP, show that spin polarized CrO has high spin (4/2) and thereby one electron occupies the doubly degenerate $e_g$ band. Therefore, because of the local moment formation, the fourth electron in CrO goes into an $e_g$ state. This strong magnetoelastic effect it is expected to be accompanied by large fluctuations of local charge distribution[27], putting the CrO system in an unstable situation against a spontaneous charge disproportionation reaction. In fact, all attempts to synthesize CrO finished with Cr + $Cr_2O_3$. Therefore, we propose that the inaccessibility of CrO at ambient pressure might be related to correlation driven electronic phase segregation[28].

Our main findings are incorporated in a new phase diagram for the monoxides of the first transition series, presented in Figure 7. The $T_C$ fro $TiO_x$ is from ref.[12]. We claim that the origin of the behaviour of VO is related to the tendency of the *d* electrons to form local spin moments in the vicinity of a metal-insulator transition. From the insulating side, this transition has been recently observed in MnO under hydrostatic pressure[29] which changes the U/W ratio keeping the number of electrons constant. Our data on VO shed light on the behaviour of the metallic side approaching the localized limit upon doping. Interestingly, the compound at which the metal insulator transition is expected, CrO, is not stable.

In summary, based upon careful thermodynamic and transport experiments on a variety of VO and TiO samples we have presented a global picture of the 3d transition metal monoxides and their metal-insulator transition. We have presented compelling evidence to claim that, in spite of its simple chemical and crystallographic structure, VO is a correlated metal with an exotic electronic phenomenology, similar to other strongly correlated systems. We also hope our results will stimulate an experimental confirmation of the pseudogap by direct spectroscopic measurements.

**Acknowledgements.** M. C. Aronson, G. Kotliar, D. Khomskii, L. H. Tjeng, S. S. Saxena, and L. E. Hueso, are acknowledged by discussion and critical reading of the manuscript. H.-D. Zhou is acknowledge by assistance during transport measurements. We also thank financial support from Xunta de Galicia (Project PXIB20919PR) from the Ministry of Science of Spain for support under MAT2004-05130-C02-01, MAT2005-06024-C02-01 and Program Ramón y Cajal (F.R.).